\documentclass[manuscript]{aastex631}  
\usepackage{booktabs}
\usepackage{multirow}
\usepackage{graphicx}
\usepackage{subfigure}

\begin{document}

\title{KIC 10855535:An elegant $\delta$ Scuti pulsator with Amplitude and Phase Modulation}

\author[0009-0003-3618-4841]{Li-xian Shen}
\affiliation{Xinjiang Astronomical Observatory,Chinese Academy of Sciences, Urumqi,Xinjiang 830011,People's Republic of China}
\affiliation{School of Astronomy and Space Science, University of Chinese Academy of Sciences, Beijing 100049, People’s Republic of China}
\author[0000-0003-1845-4900]{Ali Esamdin}
\correspondingauthor{Ali Esamdin}
\email{aliyi@xao.ac.cn}
\affiliation{Xinjiang Astronomical Observatory,Chinese Academy of Sciences, Urumqi,Xinjiang 830011,People's Republic of China}
\affiliation{School of Astronomy and Space Science, University of Chinese Academy of Sciences, Beijing 100049, People’s Republic of China}
\author[0000-0001-6354-1646]{Cheng-long Lv}
\affiliation{Xinjiang Astronomical Observatory,Chinese Academy of Sciences, Urumqi,Xinjiang 830011,People's Republic of China}
\author[0000-0002-2463-4943]{Hao-zhi Wang}
\affiliation{Xinjiang Astronomical Observatory,Chinese Academy of Sciences, Urumqi,Xinjiang 830011,People's Republic of China}
\affiliation{School of Astronomy and Space Science, University of Chinese Academy of Sciences, Beijing 100049, People’s Republic of China}
\author[0000-0002-1859-4949]{Tao-zhi Yang}
\affiliation{Ministry of Education Key Laboratory for Nonequilibrium Synthesis and Modulation of Condensed Matter, School of Physics, Xi'an Jiaotong University, Xi'an 710049, People's Republic of China}
\author{Rivkat Karimov}
\affiliation{Ulugh Beg Astronomical Institute, Uzbekistan Academy of Sciences, Astronomicheskaya 33, Tashkent, 100052, Uzbekistan}
\author{Shuhrat A. Ehgamberdiev}
\affiliation{Ulugh Beg Astronomical Institute, Uzbekistan Academy of Sciences, Astronomicheskaya 33, Tashkent, 100052, Uzbekistan}
\affiliation{National University of Uzbekistan, Tashkent, 100052,Uzbekistan}
\author{Hu-biao Niu}
\affiliation{Xinjiang Astronomical Observatory,Chinese Academy of Sciences, Urumqi,Xinjiang 830011,People's Republic of China}
\author{Jin-zhong Liu}
\affiliation{Xinjiang Astronomical Observatory,Chinese Academy of Sciences, Urumqi,Xinjiang 830011,People's Republic of China}
\affiliation{School of Astronomy and Space Science, University of Chinese Academy of Sciences, Beijing 100049, People’s Republic of China}

\begin{abstract}
We investigated the pulsating behavior of KIC 10855535 using Kepler 4-year long cadence data. Two independent frequencies were detected: a pulsation frequency $\text{F}0 = 17.733260(5){\text{d}^{-1}}$ and a low frequency $f_{8}=0.412643(8){\text{d}^{-1}}$. We identify F0 as the fundamental frequency, at which a equidistant quintuplet is centered, suggesting that the star orbits in a binary system. The fitted orbital parameters align well with those reported in previous literature. Long-term phase modulation caused by binarity has been confirmed by considering TESS light curve. Through adjusting light time via removing the light time effect, we measured a linear change in period of order $\dot{P}/P\simeq 1.44\times 10^{-7}\text{yr}^{-1}$, a value that could be indicative of stellar evolution. The star also exhibits a gradual and stable amplitude growth, thereby raising the possibility of structural changes during its evolution. We attributed$f_{8}$ and its two harmonics to rotation and surface spots, with further analysis suggesting evolving characteristics over time. Based on the hypothesis, KIC 10855535 may rotate slowly for its type, with a speed of $37(2)$km/s. Overall, KIC 10855535 presents an exceptionally clean spectrum and a relatively slow rotation as a $\delta $ Sct pulsator, exhibiting a single pulsation mode that undergoes both amplitude and phase modulation. 
\end{abstract}


\section{Introduction} \label{sec:intro}
 $\delta $ Sct stars are famous for their intermediate masses and complicated pulsating behavior. This type of stars lie on the intersection of main sequence and classical instability strip in Hertzsprung-Russell (HR) diagram, with typical spectral types of A and early F as well as effective temperature ($T_{\text{eff}}$) ranging from 6400 to 8600 K \citep{2000ASPC..210....3B,2011A&A...534A.125U,2018MNRAS.476.3169B}. $\delta$ Sct stars oscillate in p-modes, driven by the $\kappa$ mechanism  which operates in the He \uppercase\expandafter{\romannumeral2} ionization zone, where the process is analogous to heat engine cycle\citep{2000A&A...360..603T,2021RvMP...93a5001A}. These stars also oscillate in g-modes where turbulent pressure in H/He \uppercase\expandafter{\romannumeral1}  ionization layer contributes to the restoration of perturbations \citep{2000ASPC..210..454H,2014ApJ...796..118A,2019MNRAS.490.4040A}. Recently,  the edge-bump mechanism has also been found contributing to pulsation driving in Am/Fm stars\citep{1979ApJ...227..935S,2020MNRAS.498.4272M}. Typical periods for p-mode pulsations in $\delta$ Sct stars range from 15 min to 5 hr\citep{2011A&A...534A.125U,2022afas.confE...1K}. Asteroseismic research on $\delta$ Sct stars usually needs a crucial and tough step called mode identification due to their complicated pulsation modes \citep{2020Natur.581..147B,2021RvMP...93a5001A}.                           

There is a subclass of these stars, known as high-amplitude $\delta$ Sct (HADS) stars, exhibiting peak-to-peak light amplitudes exceeding 0.3 mag and being accompanied by projected rotation velocities ($v\sin i$) lower than 30 km/s\citep{2000ASPC..210..373M}. Their periods fall within the range of 1-6 hr. Majority of HADS are metal-rich and belong to Population \uppercase\expandafter{\romannumeral1}. In contrast, their counterparts in Population \uppercase\expandafter{\romannumeral2}, identified as SX Phe stars, have lower metallicity than typical $\delta$ Sct stars but can still be considered relatively metal-rich among 'halo' A-type stars.\citep{2000ASPC..210....3B,2012MNRAS.426.2413B,2017MNRAS.466.1290N,2019MNRAS.490.4040A}. In analysis on SX Phe candidates in Kepler field, at least one-third of them are found in binary (or triple) systems\citep{2017MNRAS.466.1290N}. There is a common feature in HADS that they own few identified modes. In most cases, only one or two pulsation modes will be found with most of them are radial ones\citep{2019ApJ...879...59Y,2023ApJ...943L...7L}. In recent years, even with great advancement in observing devices,  few HADS show complicated pulsations\citep{2018ApJ...863..195Y,2021MNRAS.504.4039B}. It is still a question that what exactly  determines the number of radial modes existed in pulsating stars\citep{2022ApJ...932...42L}.

To better study pulsators with subtle brightness changes, asteroseismologists can take advantage of the high-precision photometric observations provided by space telescopes like the Kepler Space Telescope\citep{2010Sci...327..977B} and the Transiting Exoplanet Survey Satellite (TESS, \citealt{2015JATIS...1a4003R}), although they were primarily designed for exoplanet exploration.  Unprecedented opportunities have been offered to explore stellar interior and thus to expand asteroseismological field since their launches.

Besides the light variation caused by periodic pulsation, irregular patterns in light curves draw more attention due to their underlying physical causes. There are two distinct types of modulation phenomena that exhibit perturbance mixing with pulsation: amplitude and phase modulation.  Some $\delta$ Sct stars show characteristics of both of them while some only have one or none of them.  

Amplitude modulation(AMod) is quite common in $\delta$ Sct stars. A survey by \citet{2016MNRAS.460.1970B} found that $61\% $ of $\delta$ Sct stars observed by Kepler exhibit at least one pulsation mode that varies significantly in amplitude over a 4-year period. The causes of this modulation can be divided into intrinsic and extrinsic ones. Intrinsic causes include beating by close frequencies, non-linearity  and mode coupling while extrinsic ones refer to binary or multiple star systems.  

Regarding phase modulation, it is a manifestation of frequency modulation(FM), in case of binaries, orbital motion induces light travel time variation and results in the splitting of pulsation frequencies into multiplets in Fourier space due to the Doppler effect\citep{2012MNRAS.422..738S,2015MNRAS.450.3999S}. It enables us to derive orbital information from frequency multiplets.  Once the frequency is fixed, phase modulation will show up instead. \citet{2014MNRAS.441.2515M} successfully presented the phase modulation method for detecting hidden binaries including variable stars, demonstrating its effectiveness in calculating orbital parameters and even radial velocities through period analysis. This method has become an important tool for supplementing studies of binary systems\citep{2018MNRAS.474.4322M}. For instance, by combining Kepler and TESS datasets, it is more likely to find non-eclipsing binaries with long orbital periods\citep{2023ApJ...943L...7L}. In addition to this, internal physical mechanisms also contribute to phase modulation, such as rotation splitting pulsation frequencies\citep{Murphy2015}, the interaction  involving energy exchanging between different pulsation modes \citep{2014ApJ...783...89B} and stellar evolution directly affecting pulsating periods\citep{2021MNRAS.504.4039B}.

Studying these modulations and their origins opens new avenues for understanding stellar interiors and their environments. A best-studied case is the evolved $\delta $ Sct variable 4 CVn, various pulsation modes exist in this star and most of them show systematic significant period and amplitude changes on a timescale of decades. Since its discovery by  \citet{1966Obs....86...34J}, observations and following study over 40 years have then begun\citep{1980LNP...125....7F,1999A&A...349..225B,2008CoAst.153...63B,2014A&A...570A..33S}. Unpredicted appearance and disappearance of modes have been illustrated by amplitude variation of this star. There was a dramatic decrease in amplitude of a single p mode at $\nu =7.735{\text{d}}^{-1}$  from 15 mmag in 1974 to 4 mmag in 1976 and to 1mmag in 1977. Subsequently a phase jump occurred accompanied by increasing amplitude, which demonstrating the growth of another new mode\citep{2000MNRAS.313..129B}. Additionally, two pulsation mode frequencies constituting beating pairs, $6.1007 {\text{d}}^{-1} $ and $ 6.1170{\text{d}}^{-1}$ , were highly variable in amplitude over the observations. By utilizing these changes, models can be well reconstructed to match observations\citep{2009AIPC.1170..410B}.
The latest related work by \citet{2017A&A...599A.116B} ascribed period changes to redistribution of angular momentum, strongly influenced by rotation due to a correlation between the direction of the period changes and the identified azimuthal order, m.

KIC 8054146 serves as a good example of resonant mode coupling theory explaining amplitude variation in $\delta$ Sct stars. The hypothesis proposed that two parent modes driven by $\kappa$ mechanism nonlinearly excited a daughter mode and meanwhile the amplitudes and phases of this triplet conform to fixed relationships respectively\citep{2014ApJ...783...89B,2023ApJ...950....6M}. However, not all cases can be explained by non-linearity as the definitions of parent and daughter modes sometimes are too strict to be applied. When studying decrease of amplitude in one single mode of KIC 7106205, \citet{2014MNRAS.444.1909B} excluded possibilities of non-linearity  and raised another question of whether energy conserved in visible pulsation modes or in other words whether mode coupling changes amplitude significantly, by investigating the star in detail, they finally provided a negative answer nevertheless. 

There seems no unified and comprehensive understanding for all modulation phenomena in $\delta$ Sct stars, but to expand our samples and acquire diverse knowledge of these intermediate-mass main-sequence variable stars, deep study on various causes is still worth it.

KIC 10855535 ($\alpha_{2000}=19^h:16^m:5^s.8$, $\delta_{2000}=+48^{\circ} :12^{\prime}:11^{\prime\prime}.3$) is classified as a $\delta$ Sct pulsator in a binary system by \citet{2018MNRAS.474.4322M}. They applied the phase modulation (PM) technique and determined that it is a single pulsator in a binary system with an orbital period of $411.92\pm0.38 $days. The other orbital parameters were calculated as $a_1\sin i /c=143.47^{+0.85}_{-0.83}$,  $e=0.0930\pm 0.012$ and $f(M)=1.8680 \times 10^{-2}\pm0.00033$ $M_\odot$. Previously, this star was studied as an overcontact binary rather than a $\delta$ Sct star and was included in the Kepler Eclipsing Binary Catalog (KEBC)\footnote{\url{http://keplerebs.villanova.edu/}}\citep{2011AJ....142..160S,2016AJ....151...68K}. \citet{2014AJ....147...45C} studied KIC 10855535 as a pair of close binaries and listed it in a candidate three-body system by analysis on precise eclipse times. Through a comprehensive study on eclipse time variation (ETV), \citet{2016MNRAS.455.4136B} suggested KIC 10855535 is a false positive binary effected by a third body and its modulations on light curves are most probably due to the pulsations of a single star instead of an eclipsing binary (EB) or ellipsoidal variables(ELV). \citet{2018MNRAS.479..183B} presented a catalogue of $\delta$ Sct which includes KIC 10855535. In brief, early studies primarily referenced the KEBC, while later studies began categorizing this star as a $\delta$ Sct pulsator in respective catalogues. This paper, however, conducts a detailed study on the target: KIC 10855535. 

The structure of this paper is as follows: Section \ref{sec:observations} covers the observation and data reduction of KIC 10855535. Section \ref{sec:frequency} focus on the analysis of its frequency properties. The phase, amplitude and rotation modulation of KIC 10855535 are discussed in Section \ref{sec:PM},\ref{sec:AM} and \ref{sec:low fre}, respectively. Finally, discussion and a summary are provided in Section \ref{sec:summary}.

\section{Observations and Data Reduction} \label{sec:observations}

\begin{figure}[ht!]
	\plotone{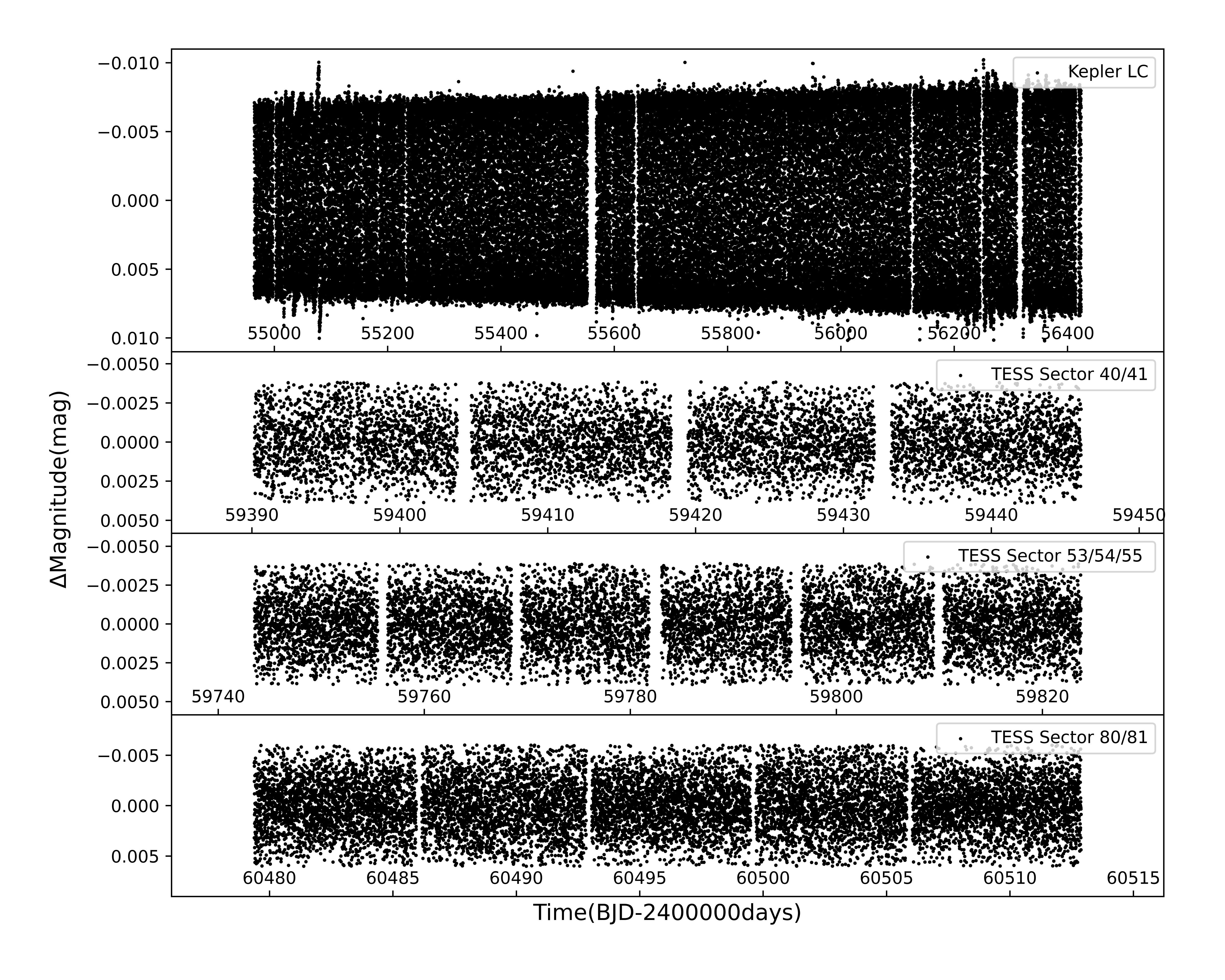}
	\caption{Light curves of KIC 10855535. From Kepler LC data, the amplitude of lightcurve is about 7mmag at the start time and increases to approximately 8 mmag at the end of observation. The sampling frequency varies in different sectors of TESS, resulting in varying densities of data points. 
		\label{fig:lightcurve}}
\end{figure}

\begin{figure}[ht!]
	\plotone{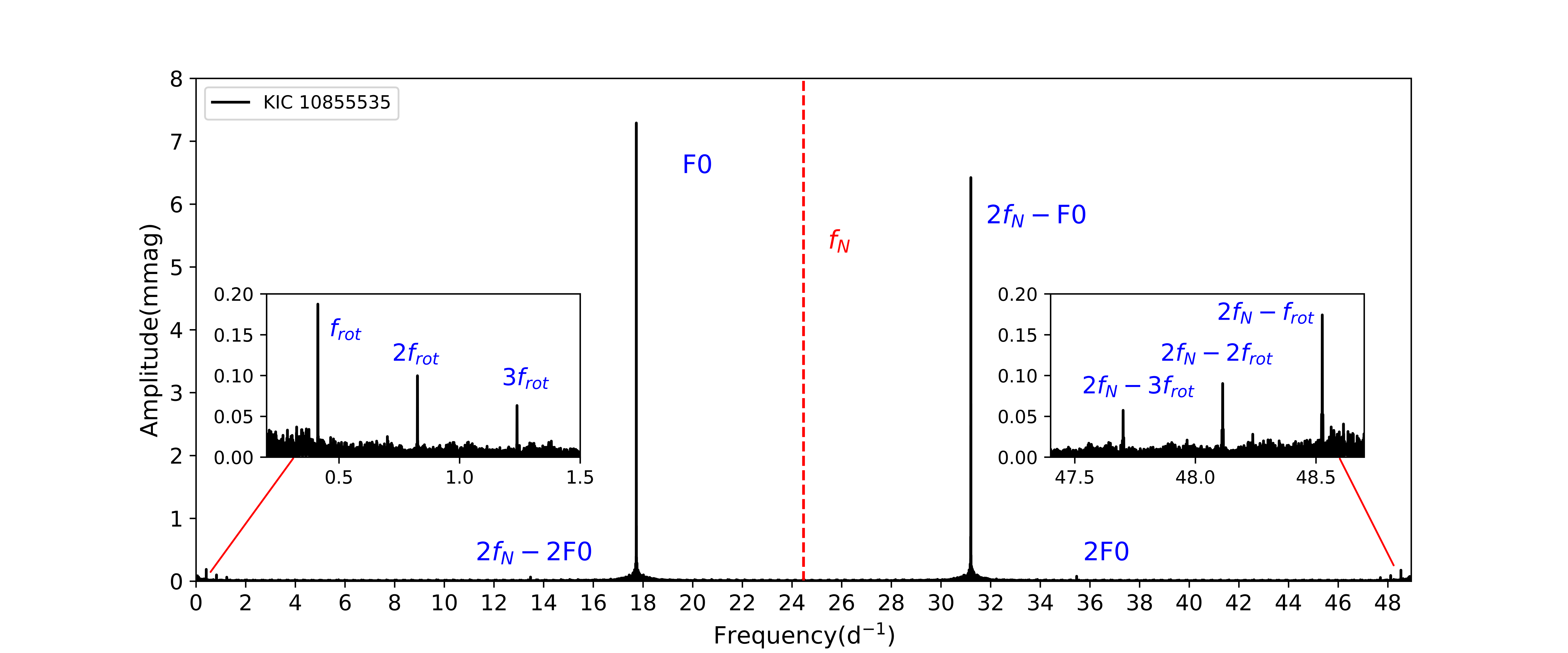}
	\caption{The amplitude spectrum of KIC 10855535 derived from Kepler LC data. Red vertical dashed line represents the Nyquist frequency($f_N=24.47{\text{d}}^{-1} $). The low frequency region is zoomed for better inspection. Note that the group of mirror frequencies (symmetrical about the Nyquist frequency ) of low frequencies have too low SNR to be listed in Table\ref{tab:frequency} but are clear in the spectrum.}
	\label{fig:spectrum}
\end{figure}

\begin{figure}[ht!]
	\plotone{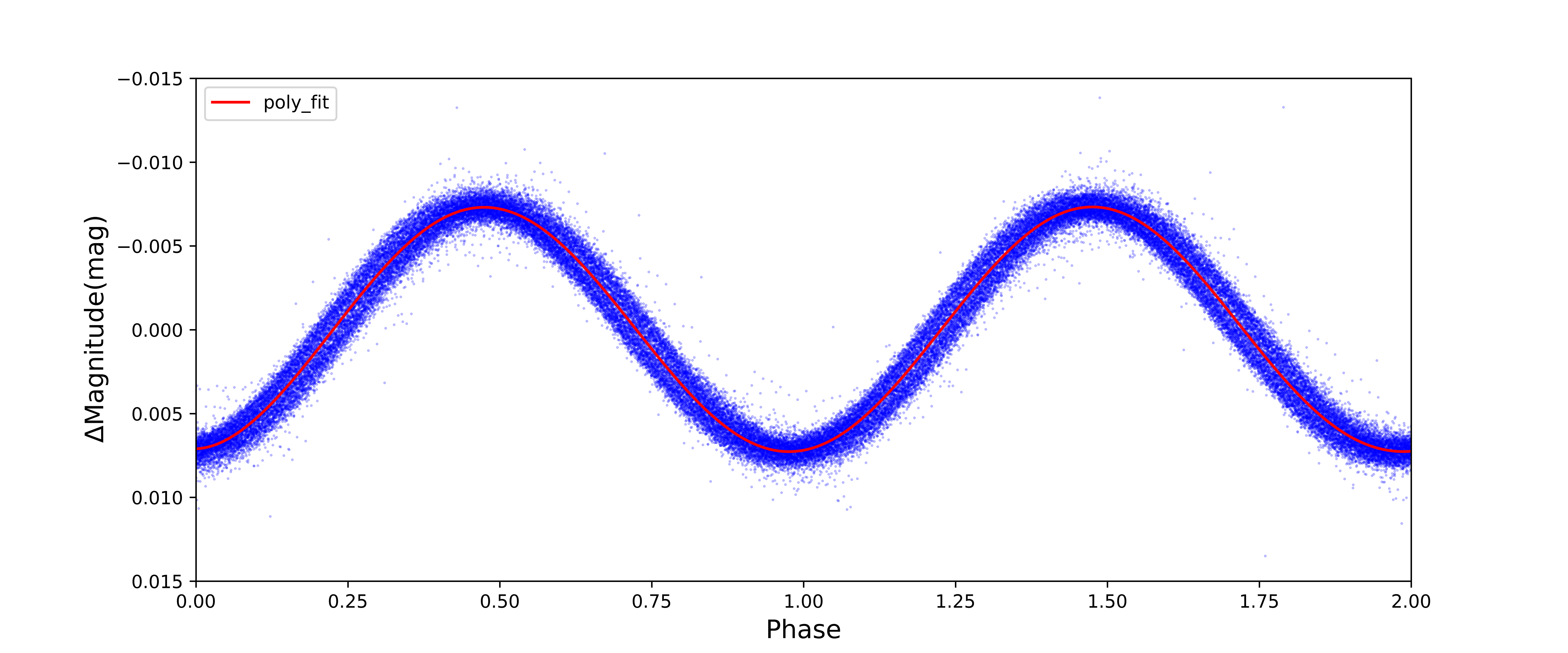}
	\caption{Phase diagram of KIC 10855535 from Kepler data, folded by F0.  The red line represents the polynomial fit for sliding average result with phase bin of 0.001 and sliding gap of 0.0001. For better clarity, two periods are plotted. }
	\label{fig:phase}
\end{figure}

\begin{figure}[ht!]
	
	\plotone{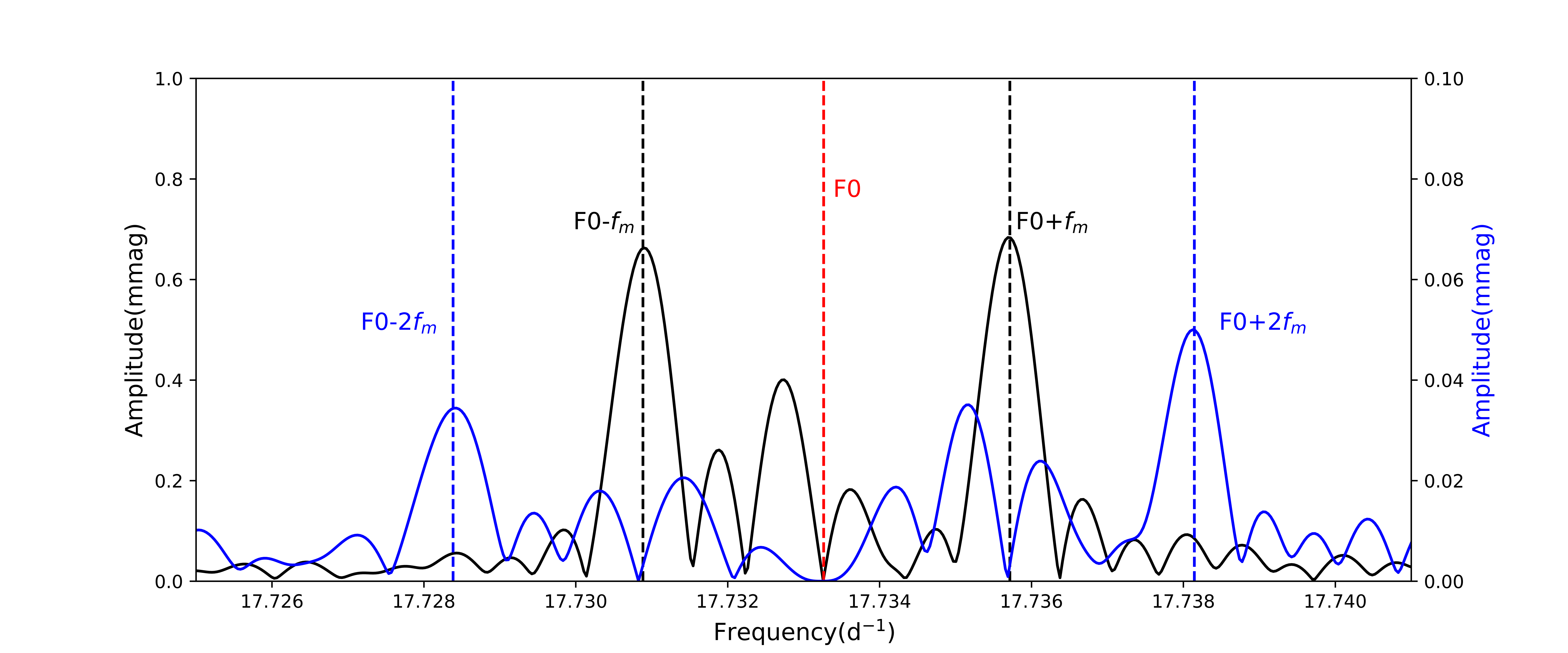}
	\caption{An equidistant quintuplet centered on the peak at $17.733260{\text{d}}^{-1}$. The black solid curve represents amplitude spectrum of KIC 10855535 after prewhitening F0, while the blue one represents the spectrum after removing all other frequencies in p-mode region(  $\geq5{\text{d}}^{-1}$). All vertical dashed lines mark precise position of quintuplets. Note different y-axis scales for blue and black lines, they have been scaled for easier comparison.}	
	\label{fig:quintuplet}   
\end{figure}

KIC 10855535 was observed by Kepler space telescope from BJD 2454964.512 to 2456424.001, spanning over about 1459 days. The data includes 17 quarters(i.e., Q1-Q17), only long cadence(LC) data (29.5 minute integration time) are available on Kepler Asteroseismic Science Operations Center (KASOC) database \citep{https://doi.org/10.1002/asna.201011437}  \footnote{\url{https://kasoc.phys.au.dk/} The website is not being maintained, but the data used in this work was downloaded before its cessation of maintenance.}. The time series data we use was provided by the KASOC Working Group 4 (WG \# 4: $\delta$ Sct targets) in which corrected flux was directly obtained from the pipeline. We then performed corrections on it, including removing some linear trends, eliminating outliers, converting photometric flux to magnitude scale and subtracting the mean value from each quarter. Finally, all the quarters were stitched to a total light curve with 65418 data points. The rectified lightcurve is shown in the top panel of Figure \ref{fig:lightcurve}, where the increasing trend in amplitude is evident from visual inspection.

Up to now, TESS has been observed this star in 10 sectors with different exposure time. Both observation times of Kepler and TESS are listed in Table \ref{tab:observations}. Sector 14, 74 , 75 have been excluded due to short time span and unknown pollution in raw files. Only the target pixel files (TPFs) which contain images for each observation cycle are available on the Mikulski Archive for Space Telescopes (MAST) \footnote{\url{https://mast.stsci.edu/portal/Mashup/Clients/Mast/Portal.html}}. From there we downloaded the rawest 15$\times$15 squares pixel data centered at KIC 10855535, and applied simple aperture photometry (SAP) method using $Lightkurve$ software\citep{2018ascl.soft12013L}. After setting threshold to confine target pixels and subtracting background from a stack of images, we can roughly obtain time-series flux variation of this star. However, due to faintness of this star(Kmag=13.87mag) and poor accuracy in data accessed through this simple method, this paper only makes use of these data in a limited way.

\begin{table}[htbp]
    \centering
    \caption{The detailed information of the Kepler and TESS Observation of KIC 10855535.}
    \label{tab:observations}
    \begin{tabular}{cccccc}
        \toprule
        Mission & Cadence & Start Time(BJD) & End Time(BJD)& Cadence(min) & Time Span (days) \\
        \midrule
        Kepler & Q1 - Q17 & 54964.0 (2009-05-13) &56423.5 (2013-05-11)& 29.4 & 1459.5\\
        TESS & Sector 40-41& 59390.2 (2021-06-25) &59418.4 (2021-07-23)& 10 & 54.8 \\
        TESS & Sector 53-55 & 59743.5 (2022-06-13)  &59823.8 (2022-09-01)& 10& 78.4\\
        TESS & Sector 80-81(partial)& 60479.4(2024-06-18)&60512.9(2024-07-21)& 3 & 33.5\\
        \bottomrule
    \end{tabular}
\end{table}

\section{Fourier analysis}\label{sec:frequency}
We performed Fourier transform on Kepler LC data through PERIOD04 \citep{2005CoAst.146...53L} , and the light curve is fitted with the following formula:
\begin{equation}
    m= m_0+\mathop{\Sigma}\limits_{i} A_i\sin(2\pi(f_i t+\phi_i))
    \label{sinfunc}
\end{equation}
where $m_0$ denotes the zero-point of magnitude, $A_i,f_i,\phi_i$ denotes the amplitude, frequency and corresponding phase, respectively.

The amplitude spectrum of KIC 10855535 appears clean, as illustrated in Figure \ref{fig:spectrum}, with a prominent peak and noticeable comb-like low frequencies.  Detailed extracted frequencies are listed in Table \ref{tab:frequency}. We adopted a criterion of signal-to-noise ratio (S/N) $\geq$ 5.0, as suggested by \citet{10.1093/mnrasl/slu194}, to identify significant peaks. Frequencies lower than 0.2 ${\text{d}}^{-1}$ were excluded due to irregular gaps in the data and various instrumental noise. Using a frequency resolution of 0.001 ${\text{d}}^{-1}$, estimated following \citep{Loumos1978} with 1.5/T, one frequency ($f_c=17.733369(2){\text{d}}^{-1})$ that was too close to F0 was also excluded. Initially, the frequency range of the Fourier spectrum was set to $0-2f_N$. Besides $f_2$ (alias) and $f_3$ (harmonic), no significant peaks were recognized. To avoid the effects of frequency aliasing, the range was then adjusted back to $0-f_N$, within which the quintuplet can be clearly detected. After applying these three filtering conditions and two separate steps, only 10 frequencies remained. The amplitude spectrum is shown in Figure \ref{fig:spectrum}.

\begin{table}[htbp]
    \centering
    \caption{Frequencies in Kepler LC data of KIC 10855535}
    \label{tab:frequency}
    \begin{tabular}{ccccc}
        \toprule
        $f_i$ & Frequency (day$^{-1}$) & Amplitude (mmag) & S/N & Comments \\
        \midrule
        1 & 17.733260(5)& 7.29(4)& 1399.2& F0 \\
        2& 31.203355(2)& 0.72(4)& 173.8&2$f_{\text{N}}-$F0\\
        3& 35.406533(5)& 0.08(4)& 42.4&2F0\\
        4& 17.735681(4) & 0.65(4) & 167.9 & F0$+f_m$ \\
        5& 17.730871(3)& 0.66(4)& 239.0& F0$-f_m$ \\
        6& 17.738134(2) & 0.05(4) & 26.5& F0$+2f_m$ \\
        7& 17.728398(5) & 0.04(4) & 18.9& F0$-2f_m$ \\
        8& 0.412643(8)& 0.19(4)& 23.4&$f_{\text{rot}}$ \\
        9& 0.825553(7)& 0.10(4) & 14.5& $2f_{\text{rot}}$ \\
        10& 1.238240(5)& 0.06(4) & 13.7& $3f_{\text{rot}}$ \\
        \bottomrule
    \end{tabular}
    \tablecomments{Two frequenies ($f_1, f_8$) among these are independent ones, others are harmonics or combinations. F0, $f_m$,$f_N$ and $f_{\text{rot}}$ represents fundamental, modulation, Nyquist and rotation frequency, respectively. }
\end{table}

Figure \ref{fig:phase} shows phase-folded light curve at  $\nu=17.73326{\text{d}}^{-1}$ which denotes the highest peak in amplitude spectrum(Figure \ref{fig:spectrum}).  The misunderstanding, claiming that KIC 10855535 is an overcontact binary, arose because its phase-folded diagram resembles that of an EW-type binary. Based on visual inspection of its light curve and consideration of its ultra-short period, it is more logical to classify it as a variable star rather than an overcontact system. Acknowledging this classification, we proceed to study its photometric traits.

The Nyquist frequency of LC data is $f_N = 24.469{\text{d}}^{-1}$. It is not difficult to find $f_2=31.203355{\text{d}}^{-1}$ is just a combination frequency of $f_N$ and $f_1$, and $f_3=35.466533{\text{d}}^{-1}$ is the harmonic of $f_1$. Other frequencies can be divided into two groups based on typical pulsating frequency of $\delta$ Sct \citep{2011A&A...534A.125U}. A group of equidistant quintuplet shown in Figure \ref{fig:quintuplet}, which is separated by $f_m= 0.00243(1){\text{d}}^{-1}$, was found within $\delta$ Sct regime ($\nu \geq 5{d}^{-1}$). Another three peaks in lower region can been seen as $f_8$ and its harmonics. Following analyses will only focus on $f_1$, $f_8$ and $f_m$.
$f_1$ was identified as fundamental frequency. We first assumed this statement based on common judgment as its amplitude is the largest one, then we calculated its pulsation constant Q which can be estimated by following fomula \citep{1975ApJ...200..343B}:
\begin{equation}
	\log Q = \log P +\frac{1}{2} \log g +\frac{1}{10} M_{bol} + \log T_{\text{eff}} -6.454
\end{equation}
where $P$ denotes the pulsation period, $g$ represents surface gravity, $T_{\text{eff}}$ represents effective temperature. $M_{\text{bol}}$ is the absolute bolometric magnitude whose value can be determined by:
\begin{equation}
M_{\text{bol} }= m_V +5-5\log d + BC
\end{equation}
where $m_V$ represents apparent magnitude in V band, therefore $m_V +5 - 5\log d$ denotes absolute magnitude(i.e., $M_V$) in which $d$ represents distance. BC is Bolometric Correction which can be estimated by empirical relation dependent on its effective temperature\citep{1984ApJ...284..565H}. The fundamental parameters of KIC 10855535 are listed in Table \ref{tab:parameters}. Q value of $f_1$ is derived as 0.0342(5), indicating $f_1$ corresponds to fundamental mode in this pulsating star according to \citet{2017ApJ...849...38L}, while this value contains uncertainty and insufficient to confirm $f_1$ as the fundamental radial mode, to further investigate, we conducted another check using Period-Luminosity (P-L) relation.  Different modes in $\delta$ Sct stars follows different P-L relation and it in turn could be applied for mode identification. First we conducted Spectra Energy Distribution(SED) fitting using PySSED, utilizing data from 2MASS, Gaia, Pan-STARRS and WISE \citep{2024RASTI...3...89M}, yielding  $L=9.19L_\odot$ and then compared this star's parameters with different P-L relations presented by \citet{2021PASP..133h4201P}. The result shows that $f_1$ is consistent with the fundamental mode. Considering its highest amplitude, the Q value, and its alignment with the P-L relation for the fundamental mode, we suggest that  $f_1$ corresponds to the fundamental radial mode (denoted as F0).
\begin{table}[htbp]
	\centering
	\caption{Fundamental parameters of KIC 10855535}
	\begin{tabular}{lcc} 
		\toprule
		Parameters & Values & References \\
		\midrule
		Kepler ID & 10855535 &  \\
	    TIC ID & 299157009 &  \\
		R.A.(J2000) &$19^h:16^m:5^s.8$&  \\
		Decl.(J2000) & $+48^{\circ} :12^{\prime}:11^{\prime\prime}.3$&  \\
		$Kmag$& 13.870 &(a) \\
		$m_V$ & 13.899 & (a) \\
		$\log g$ & 4.045 dex & (a)\\
		         & 4.1715 dex &  (b)\\
 & 4.124(0.038)&(d)\\
		$T_{\text{eff}}$ & 7555(200) K & (a) \\
		                 & 7271(22) K  & (b) \\
 & 7270(28) K&(d)\\
		$R/R_{\odot}$& 1.863 & (a) \\
					 & 1.741 & (b) \\
 & 1.754(0.1)&\\
		$\varpi$ & 0.539(14) mas & (c) \\
		d & 1760(42)pc & (b) \\ 
            $[\text{Fe/H}]$ & -0.285(0.022)& (d) \\
                & -0.315(0.221)& (d) \\
            \bottomrule
	\end{tabular}
	 \label{tab:parameters}
	\tablecomments{(a) Kepler Input Catalog(KIC) \citep{2009yCat.5133....0K} (b) Tess Input Catalog(TIC) \citep{2019AJ....158..138S} (c) Gaia DR3 \citep{2022yCat.1355....0G} (d) LAMOST LRS DR11 \citep{2015RAA....15.1095L} \url{ http://www.lamost.org/dr11}.} 
\end{table}

The low frequency region (0-5$\text{d}^{-1}$) in $\delta$ Sct stars could originated differently. Most cases are associated with combined frequency of other high frequencies, whilst some are independent pulsation mode (g-mode)  excited by convective flux blocking mechanism\citep{2006MmSAI..77..366D,2010ApJ...713L.192G}. In this work, we prefer to classify KIC 10855535's low frequencies as rotational effect. Given that we can only observe the peak along with its harmonics, in addition, the frequencies in high frequency region are relatively clean, it would be more reasonable to consider rotational motion rather than g mode or non-linear combination of pulsation frequencies. Then $f_8$ was denoted as $f_{rot}$.

The modulation frequency $f_m = 0.00243(1)\text{ d}^{-1}$,  is too close to zero to be easily distinguished from noise in the spectrum. It is appropriate to apply the FM technique here based on equidistant frequency separation\citep{2012MNRAS.422..738S,2015MNRAS.446.1223K}, $f_m$ represents actually the orbital frequency, illustrating a period of approximately 411.5$\pm$1.7 days.  It is consistent with the result of  $411.92\pm0.38 $ days given by \citet{2018MNRAS.474.4322M}.

\section{Phase Modulation} \label{sec:PM}
\begin{figure}	
	
	\plotone{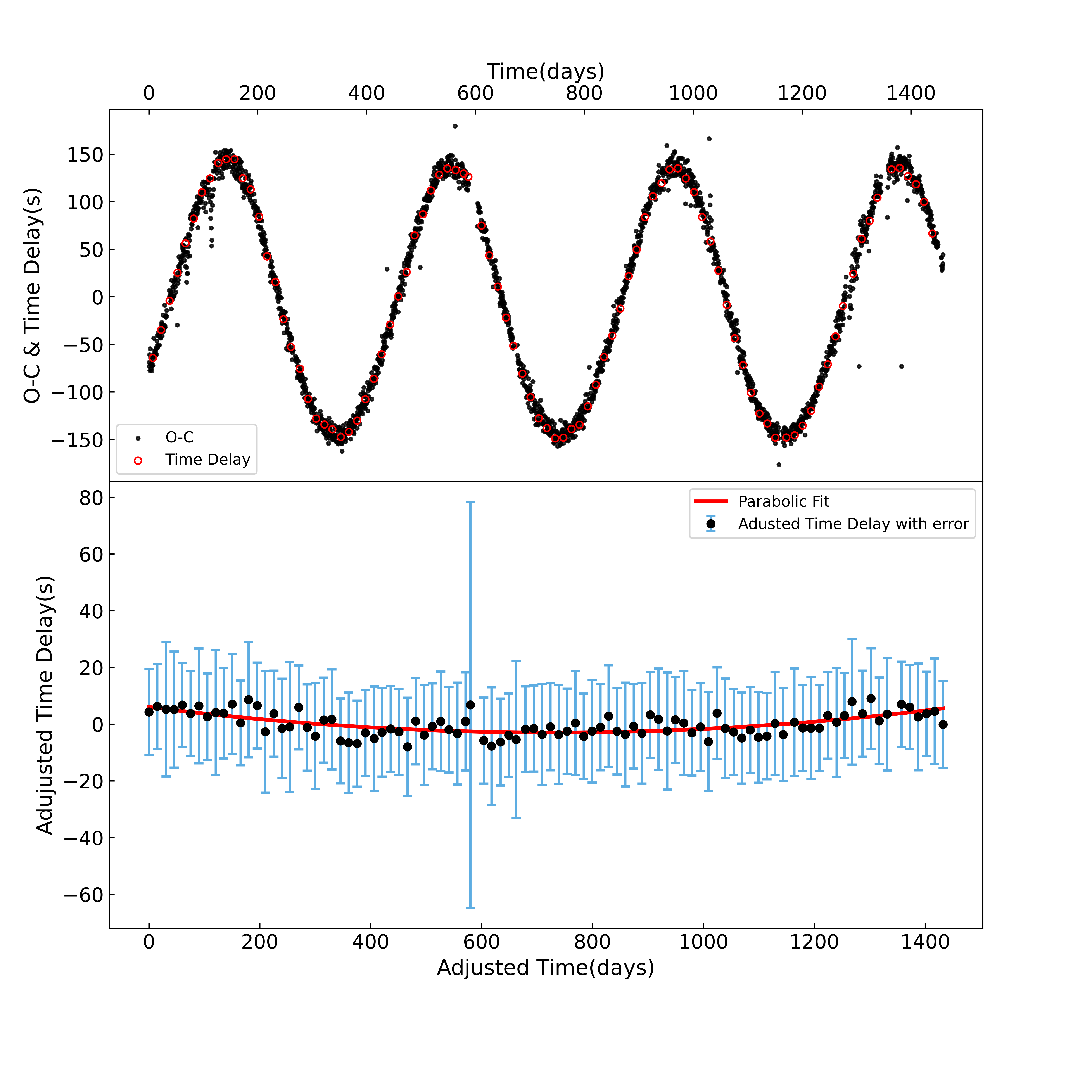}
	\caption{Phase modulation before and after removing LTE. The upper panel shows the time delay (derived from the phase modulation method) and O-C results of the F0 mode in KIC 10855535 over the Kepler observing span. Each red open circle represents the time delay measured every 15 days. The black dots indicate the O-C results, based on a period of 0.05639 days (P = 1/F0). The lower panel shows the time delays after eliminating binary effects by adjusting time axis. The red curve represents a parabolic fit . There is an abrupt error bar existing due to fewer data points in corresponding segment. }
	\label{fig:PM}
\end{figure}

\begin{figure}[ht!]
	\plotone{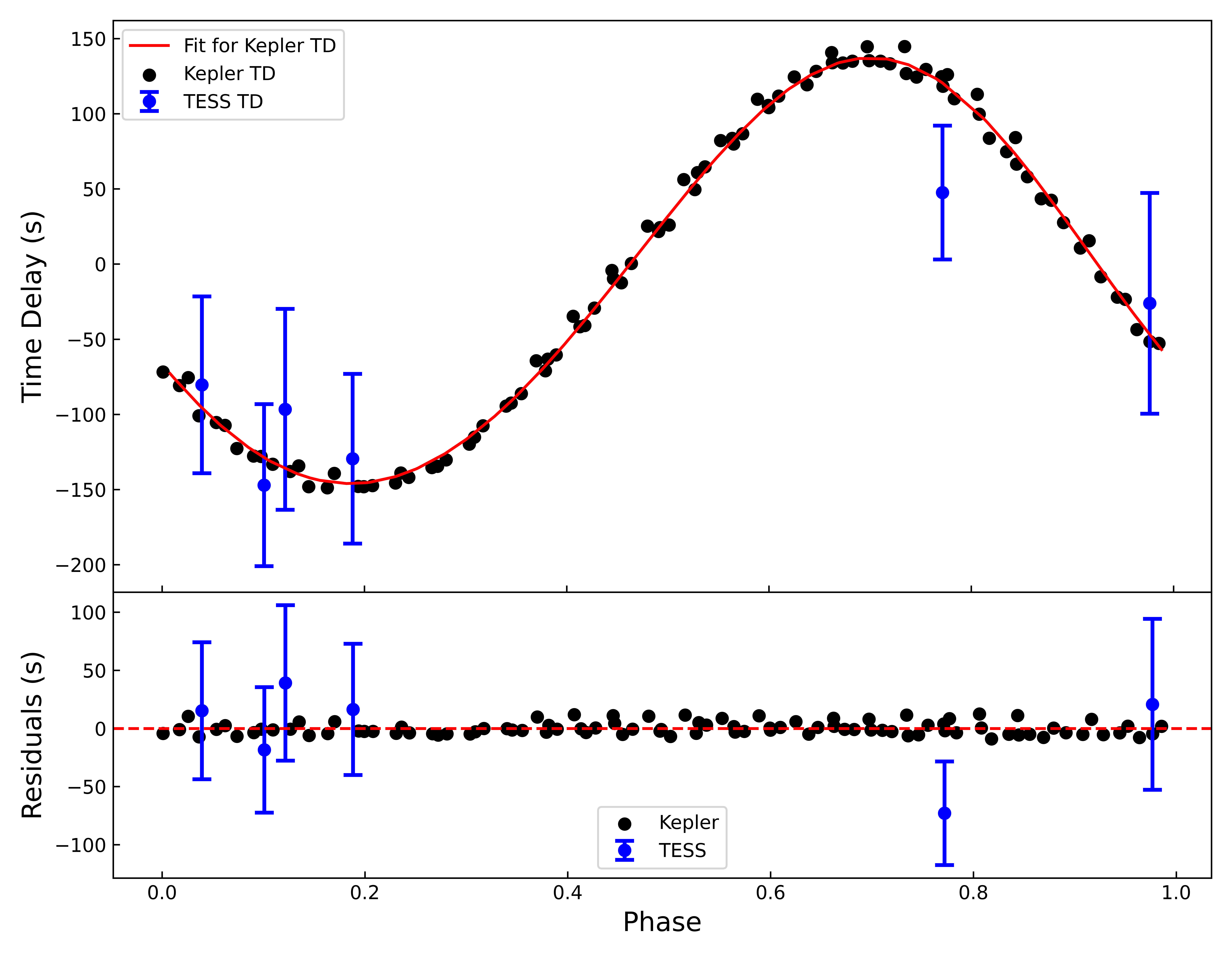}
\caption{Phased time-delay curve derived from Kepler and TESS observations at orbital frequency($0.00243{\text{d}}^{-1}$). Black and blue points represent Kepler and TESS time delays,respectively. The red curve is one of the best fit for Kepler time delays. Error bars for TESS data are derived by Bootstrap resampling. Doppler shift and constant differences between two missions have been rectified for better view.  The bottom panel shows corresponding residuals.}
	\label{fig:keplertesspm}
\end{figure}

To inspect pulsating period variation, various methods have been applied for checking phase modulation. One is 'O minus C'(O-C, \citealt{1980A&A....82..172P,2005ASPC..335....3S}), analyzing differences between observed maximum light time($T_O$) and calculated time($T_C$) which is predicted by constant period(P), cycle number(E) and the first maximum light time($t_0$) using following fomula:
\begin{equation}
	T_C = t_0 + P \times E
\end{equation}
It is noteworthy that the star pulsates with a short period, during which Kepler only sampled 2 or 3 points in LC data, making it challenging to pinpoint the moment of maximum light. To solve the problem, we sacrificed some data precision by overlapping ten periods once and utilizing clearer light curve consisting of approximately 30 points, allowing for more precise fitting to identify the brightest moment. Despite ten periods are utilized for a single light curve, we consider only one timing of maximum light within the median period. For all photometric data spanning over 1400 days, about 2400 brightest moments has been recorded through this process. Next step is calculating their corresponding cycle number and then $T_C$ along with O-C will be obtained. Another method is directly measuring the phase of the pulsation signal. By performing Fourier analysis on the pulsation signals at different time intervals , the phase values at at fixed frequency can be determined  \citep{2014MNRAS.441.2515M}.  In this work, we calculated phases of independent frequencies( F0 and $f_8$) , however only that of F0 shows significant changes with time. 

As depicted in upper panel of Figure  \ref{fig:PM}, both results exhibit quasi-sinusoidal patterns, suggesting that the pulsating period is influenced by orbital effects. Like analyzed in \citet{2018MNRAS.474.4322M}, there is no reason for us to overturn that KIC 10855535 belongs to a binary system and the periodicity of time delays shows that of orbital motion. The Light Time effect (LTE, \citealt{2005ASPC..335....3S}) can explain this in a scenario that light travels to the telescope at different phases in one orbital period, which leads to phase modulation.  Time delays($\tau(t)$) can be inferred from phase differences by this formula:
\begin{equation}
    \tau(t) = \frac{\Delta \phi(t)}{2\pi\nu},
    \label{TD}
\end{equation}
where$\Delta\phi = \phi-\bar{\phi}$ , $\bar{\phi}$ is the mean value of phases, $\nu$  denotes the fixed pulsation frequency\citep{2014MNRAS.441.2515M}. Based on the same physical background, O-C results overlap with the time delays. During first hundreds of days in Kepler observation, time delay increases and the star was moving away from us on its binary orbit.

During entire Kepler observation time, fewer than four complete orbital periods have been detected. To enhance confidence and out of curiosity, TESS observations were used to extend temporal range. To better fit the orbital parameters, Maelstrom code\citep{2016MNRAS.461.4215M,2020AJ....159..202H,2022ascl.soft05005H} which was designed to model binary orbits through the phase modulation technique have been applied on the Kepler LC data. Through forward modeling orbital binaries onto the light curve, one best-fitting model for time-delay using 15-day light curve segments is shown in red curve in Figure \ref{fig:keplertesspm}. The solution also gives other orbital parameters including an orbital period of $412.18^{+0.34}_{-0.34}$  days, a projected semimajor axis of $a_1\sin i /c = 143.07 ^{+0.78}_{-0.81}$s, eccentricity of $e=0.019^{+0.01}_{-0.01}$ and a mass function of $ f(m_1,m_2,\sin i) = \frac{(m_2\sin i)^3}{(m_1+m_2)^2}=0.09^{+0.0003}_{-0.0003}$ $M_\odot$. The parameters are consistent with those published in previous work \citep{2018MNRAS.474.4322M}. The period is also in accord with that $411.5\pm 1.7$ days derived from FM technique (Section \ref{sec:frequency}).

Figure \ref{fig:keplertesspm} shows time-delay curves folded by orbital period using both Kepler and TESS data. According to the length of data, 26 days and 30 days have been chosen to segment Sector40-41,53-55 and Sector80-81 of TESS data. The result show that they match the fit well, revealing that KIC 10855535 has been consistently orbiting around another companion for nearly 15 years(2009-2024). 

There are additional details behind this figure that deserve discussions.  TESS sectors has been divided into 3 groups by time, and 3 main frequencies around $17.73{\text{d}}^{-1}$ can be acquired, which are slightly different with the $17.73326(5) {\text{d}}^{-1}$ observed in Kepler LC data. The fixed frequency we used to calculate time delay has been fixed at $17.73326 {\text{d}}^{-1}$ for both Kepler and TESS data. The differences in frequencies are caused by observational biases, as the TESS data predominantly cover half of the orbital period when the star is closer to us. This incomplete sampling causes discrepancies in  $\tau_{\text{obs}}(t)$ and $\tau(t)$ \citep{2016MNRAS.461.4215M} and rectification can be made based on Equation \ref{dopplercorrect} \citep{2023ApJ...943L...7L}:
\begin{equation}
	\tau_{\text{obs}}(t) = \frac{1}{1+\alpha}\tau(t)+\frac{\alpha}{1+\alpha}
	\label{dopplercorrect}
\end{equation} 
where $\alpha=(\nu_{obs}-\nu)/\nu$. $\nu$ here denotes intrinsic pulsation frequency and in practical use, was replaced by F0 extracted by 4 years of Kepler LC data. 
There are also expected phase shifts in the pulsations between the Kepler and TESS data due to the different passbands used by the two missions\citep{2022ApJ...937...80M}. The values of $\Delta\phi$ are further affected by the incomplete sampling of both telescopes. As a result, we use the same constant value as a replacement for$\bar{\phi}$ in Equation \ref{TD}. We then apply a least-squares fit to adjust all points to a common baseline. The specific value of TESS time delays is meaningless and its variation tendency counts more. 

To isolate the effects of binary motion and focus on the intrinsic behavior of the pulsator, the time axis has been corrected according to the time-delay fit, restoring the light arrival times to their original values without any binary influence.  The bottom panel of Figure \ref{fig:PM} presents the recalculated time delays using the adjusted time from KLC data. There is a clear parabolic change in it and indicates a gradual period growth.  The linear period variation in pulsation modes can be determined from:

\begin{equation}
O-C = \frac{1}{2}(\frac{1}{P}\frac{dP}{dt})t^2 = \frac{1}{2}(\frac{\dot{P}}{P})t^2,
    \label{period change}
\end{equation}

where O-C equals to time delays as well,  t is the length of the observations in days, and P is the period of the signal in days. After converting all units into the conventional one of year, the fractional period change , ${\dot{P}}/{P}$ ,  can be derived as $1.44\times10^{-7} \text{yr}^{-1}$.  It is consistent with typical linear period changes of $\delta$ Sct stars, around $10^{-7} \text{yr}^{-1}$, which is at least an order of magnitude larger than those predicted by stellar evolution theory for main sequence $\delta$ Sct stars\citep{1980A&A....82..172P,1998A&A...332..958B}. While it is still unclear whether the observed changes are due to stellar evolution, the approach of studying period variations has been applied on different types of pulsators with differing levels of success\citep{1992MNRAS.254...59W,2012MNRAS.427.1418M,2021MNRAS.504.4039B}.

\section{Amplitude Modulation} \label{sec:AM}
\begin{figure}
	\centering
	\plotone{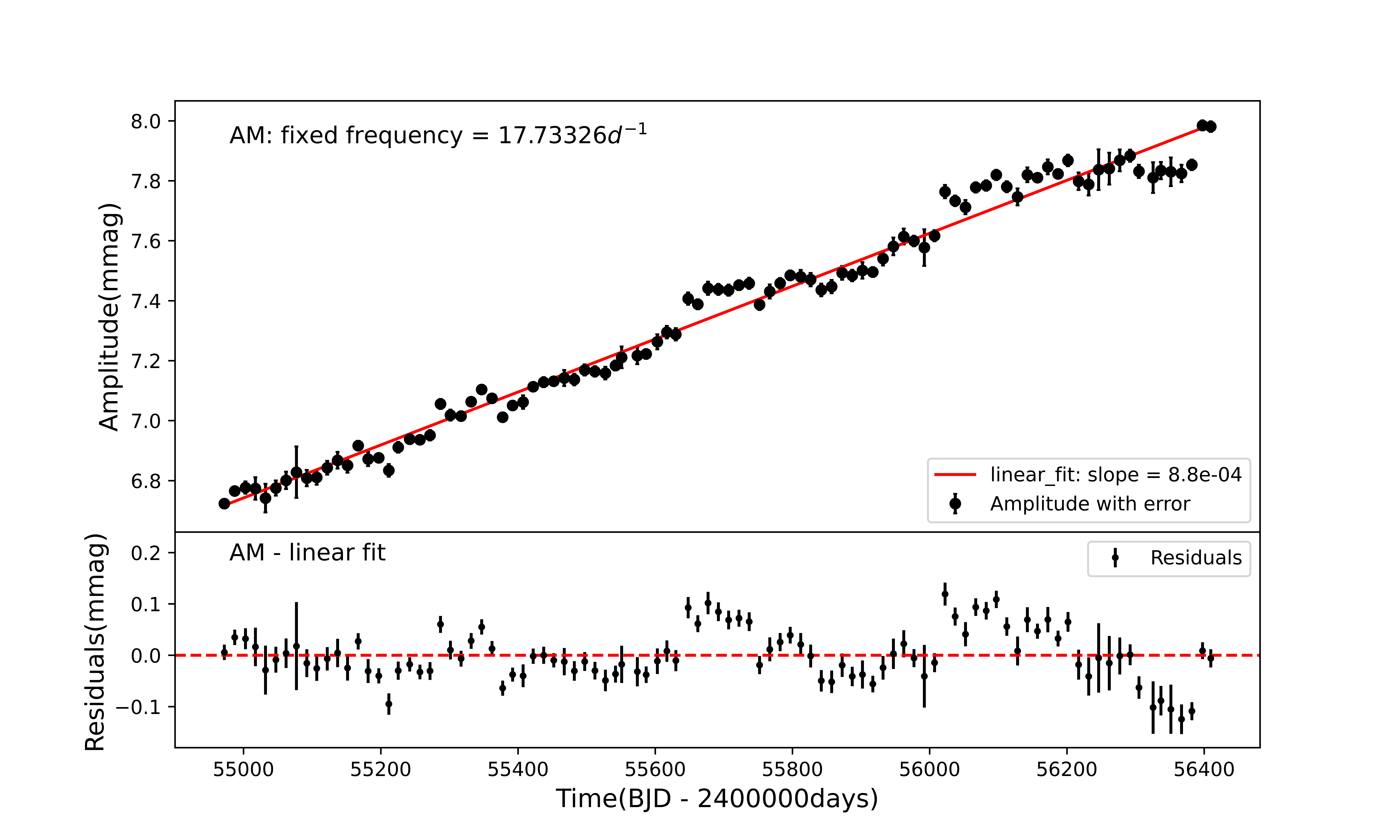}
	\caption{Amplitude modulation in F0 over 4 years. Each point in the top panel represents the fitted amplitude derived from a 15-day segment of the light curve. The red line shows a linear fitting to it.  The residuals are shown in the bottom panel.}
	\label{fig:AM}
\end{figure}

Different from phase modulation, which represents changes in period, amplitude variation reflects changes in the pulsating energy inside the star in most cases \citep{2016MNRAS.460.1970B} . To track AMod, the data has been divided into time bins with each 15 d in length (except the last bin). Amplitude and phase at fixed frequency  have been optimized  in each bin using linear least-squares. This approach has been used in work of \citet{2014MNRAS.444.1909B} and \citet{2016MNRAS.460.1970B}. The difference in this work lies in the data used. To better disentangle phase modulation effects caused by binary motion from amplitude variations, the time points have been adjusted to the light emission time based on the fitted time delays (O-C). As described in Section \ref{sec:PM}, no significant indications of binarity affecting phase modulation were found in the adjusted data. Therefore, the results shown in Figure \ref{fig:AM} can be considered unaffected by the Doppler effect in the binary system. 

\subsection{Linear Growth}
The amplitude appears to grow in a nearly linear fashion during time span of data, about +0.88 mmag per 1000d , indicating that it is steadily increasing in the mode at F0. This could be due to underlying physical processes since the timestamps have been rectified to remove the LTE. Based on previous literature working on explaining AMod, some assumptions will be discussed here.

(\romannumeral1) Mode beating 

A pair of close and resolved mode frequencies, with spacing below $0.01{\text{d}}^{-1}$, will lead to amplitude and phase modulation at the same time with the principle called beating effects which are difficult to be detected in $\delta$ Sct stars\citep{2002A&A...385..537B}. The beating behavior appears as periodic AMod, with a characteristic sharp change in phase at the epoch of minimum amplitude for each frequency in the pair\citep{2006MNRAS.368..571B}. Some of such modulations have been identified as beating like those in KIC 4641555 and KIC 8246833\citep{2016MNRAS.460.1970B}.

It is unreasonable to attribute pulsational amplitude growth to mode beating in KIC 10855535.  First of all, there is no resolved close frequencies in the amplitude spectrum, to say the least, even if not taking high precision in the order of $\mu mag$ in Kepler observations into consideration, it would be scarcely possible that another independent mode exist in such a clean spectrum while not been detected.  Secondly, AMod caused by mode beating can not be separated with phase modulation, but besides phase modulated by binary effects and gradual period growth indicating stellar evolution, there are no correspondent simultaneous variations in phase. Hence, the hypothesis is unreliable in this case.

(\romannumeral2) non-linearity interaction and mode coupling

The prerequisite for both non-linearity and mode coupling is the existence of different modes. Similar to the hypothesis of mode beating, there could be unknown pulsation modes present, though the likelihood is very small. Stars that exhibit non-linearity can be identified by the nonsinusoidal shape of their light curves and the presence of harmonics and combination frequencies in their amplitude spectra\citep{2016MNRAS.460.1970B}. Therefore, despite rather small possibility, we cannot completely exclude either possibility. 

(\romannumeral3) stellar evolution

Pulsation keeps stable with the balance of driving and damping mechanisms inside a star. The consequence of stellar evolution on the structure of a star is that an increase in stellar radius causes the period of the fundamental radial mode to increase.  Pulsation energy therefore changes on account of increasing radius of star\citep{1998A&A...332..958B}.

There are some noteworthy cases relevant to this speculation. The $\rho$ Pup star KIC 3429637, shows continuous pulsational amplitude variation during first 2-yr data of Kepler.Two dominant modes at $\nu=10.33759{\text{d}}^{-1}$ and  $\nu =12.47161{\text{d}}^{-1}$ show amplitude growth at different rates, both around +1 mmag per 1000 days. Combined with spectroscopic observations and its location near the terminal-age main sequence in evolutionary models, a hypothesis of stellar evolution has been supposed\citep{2012MNRAS.427.1418M}. However, analysis of 4 years of Kepler data reveals that this continuous growth has reversed\citep{2017ampm.book.....B}. Non-linear effects or mode coupling might also explain the non-periodic amplitude variation. Additionally, no phase modulation in its pulsation modes follows the quadratic relation described in equation  \ref{period change} which challenges expectations regarding structure changes during evolution lead to pulsation AMod, despite its position near the TAMS based on model fitting. 

Another similar case is KIC 8453431, which also owns clean frequency spetrum and contains a single pulsation mode that increases in amplitude whilst all other pulsation modes stay constant in amplitude and phase\citep{2016MNRAS.460.1970B}. It is seen as the counterpart of KIC 7106205, no definite interpretation has been given for both of their single-mode AM.  Within detectable range,  interaction between modes show no signs. 

Returning to this target KIC 10855535, it is challenging to model the evolution due to the simplicity and singularity of its pulsation modes. 
While the 4-year data is sufficient to detect period changes and amplitude variations, it is too short to draw conclusions about the star's evolutionary process. The increasing amplitude could be a flash in a pan, similar to  what occurred with KIC 3429637.  Nevertheless, KIC 10855535 differs from KIC 3429637, KIC 8453431, KIC 7106205, and other stars that show pure AMod, as expected phase modulation linked to stellar evolution has been detected in this case.  For a star on the main sequence, stellar evolution predicts an increasing period\citep{1998A&A...332..958B}, which aligns with both the phase and amplitude modulation observed  in KIC 10855535.

In conclusion, if the AMod is not a short-lived phenomenon, it suggests that structural changes due to stellar evolution are the most likely cause, even in the absence of additional evidence. Alternatively, it is also possible that AMod is unrelated to period changes altogether.

(\romannumeral4) stellar cycle

If the mode amplitude shows apparent periodic changes, then it would be the evidence of stellar cycle rather than stellar evolution\citep{2000MNRAS.313..129B,2014MNRAS.444.1909B}.
To check future tendency and verify the hypothesis of stellar cycle, we attempted to observe the amplitude variation using the TESS data for KIC 10855535, which is a faint star with a magnitude of 13.9. However,  photometric precision of TESS, on the order of milli-magnitudes, was insufficient to detect the AMod observed in Kepler, which is also in the milli-magnitude range. Consequently, no specific amplitude trend was observed in the TESS data, making this speculation still be possible.

\subsection{Fluctuation in amplitude growth}
Besides the linear increase, there appears to be a weak periodic sinusoidal signal in the residuals (the lower panel in Figure \ref{fig:AM}), which can be derived through a second Fourier transform. This time,  the main frequency is $0.002453(6){\text{d}}^{-1}$ with an S/N value of 8 and amplitude of 0.05 mmag, corresponding to a fluctuating period of 408(1) days. This becomes particularly interesting due to its proximity to the binary orbital period. The results emerged only when all significant frequencies were found. Since the LTE has been removed via adjusting light time, then the periodic fluctuation  must be caused by other effects related to binarity. Possible explanations for this could include tidal distortion, reflection, and Doppler beaming and boosting \citep{2010ApJ...715...51V,2017MNRAS.472.1538F}.  Considering that the orbital period exceeds 400 days and the orbital eccentricity is low, the influence of these factors would likely be limited. However, these effects could still manifest as amplitude variations after removing the LTE. Therefore, there must be something unique about this system, leaving the discussion open for further exploration.

\section{Low comb-like frequencies} \label{sec:low fre}

\begin{figure}
    \centering
    \plotone{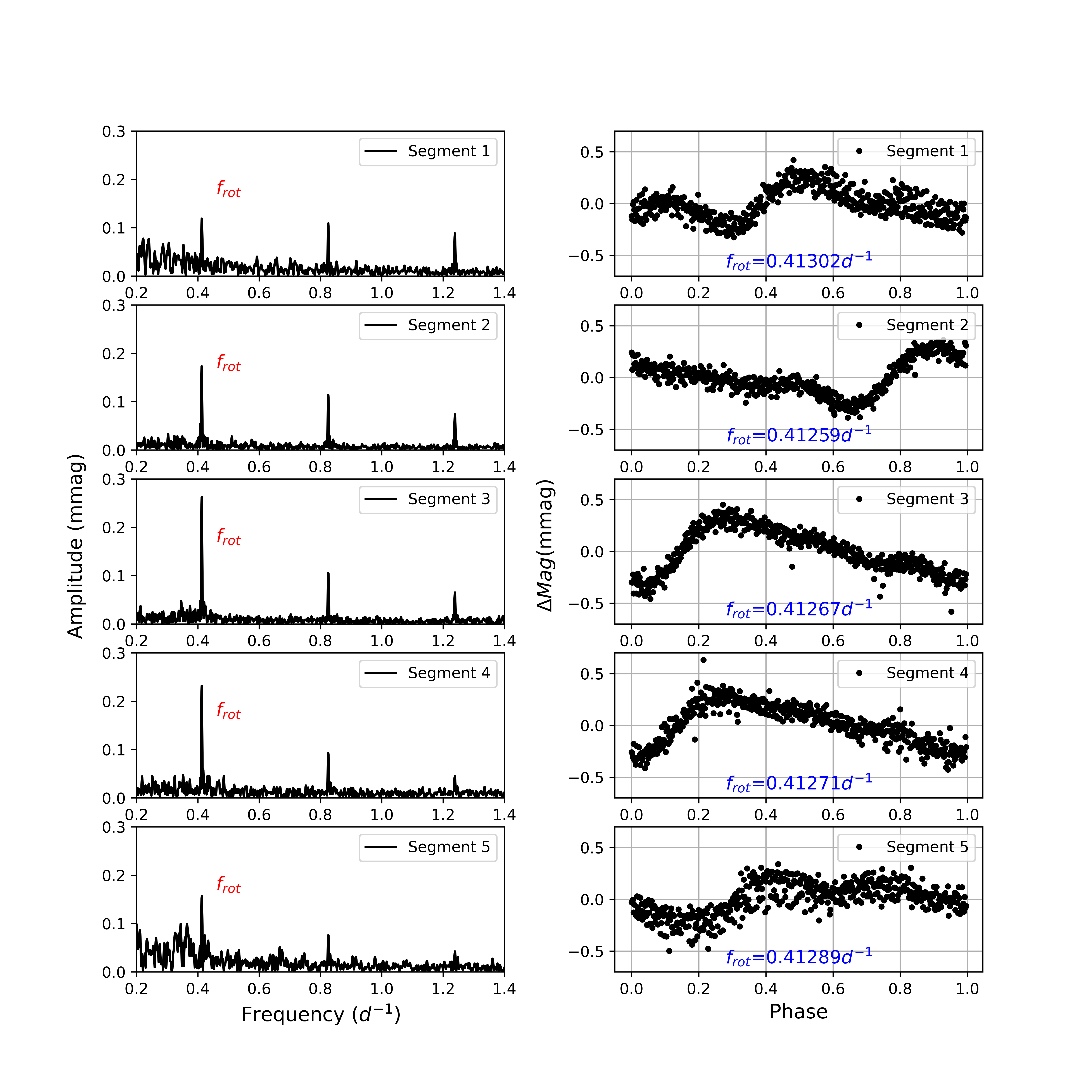}
    \
    \caption{The amplitude spectra and phase folded light curves of each segment. The phases are binned every 0.002 phase.}
    \label{fig:frot}
\end{figure}

In Section \ref{sec:frequency}, we refer the low frequency $f_8$ as rotation modulation. Here we further analyze the consequences of this attribution. To focus on the low-frequency region, we filtered out all frequencies above 5 ${\text{d}}^{-1}$ from the spectrum. We then divided the remaining data into five segments based on gaps in it, with each segment spanning approximately 300 days.  A significant gap was also removed to ensure continuity. The rotation frequency and its harmonics in each segment are listed in Table \ref{tab:lowfrequency}. With the spectrum showing on the left panel of Figure \ref{fig:frot},  the residuals of light curve are folded with the  rotational frequency and shown  in the right panel.  Not only the precise frequency peak locates slightly differently between segments, the amplitudes, phases and S/N also differ.  Based on the different shapes in phase folded diagram of each segment, these brightness variations are probably caused by star spots on the surface. 

According to \citet{2013MNRAS.431.2240B} , rotational modulation caused by magnetic activity is quite noteworthy on A-type stars while at that time no observational evidence can be found to prove that magnetic activity is common in hot stars like A-type stars.  In fact, weak magnetic field is detected in Vega, Sirius and even in a $\delta$ Sct star HD 188774\citep{2010A&A...523A..41P,2016A&A...587A.126B,2015MNRAS.454L..86N}. As to star spots,  because rotational modulation helps explain partial low frequencies appearing in hot variable stars, their existences have been confirmed in more stars like KIC 7047141, KIC 3440495, KIC 5950759 and so on \citep{2019ASPC..518..195S,2022ApJ...937...80M,2018ApJ...863..195Y}. The pattern of comb-like frequencies in KIC 10855535 is similar to those have be speculated as rotational modulation, so similarly like others,  star spots on the surface may created harmonics of $f_{rot}$ in this star.

\begin{table}[htbp]
	\caption{Frequencies extracted from each segment of residuals of Kepler LC data}
	\begin{tabular}{cccccc}
		\toprule
		\multirow{2}{*}{Segment} & \multirow{2}{*}{title} & Frequency & Amplitude& \multirow{2}{*}{Phase} & \multirow{2}{*}{S/N}\\
		&&(day$^{-1}$)&($\mu$mag)&&\\
		\midrule
		1 & $f_{rot}$ &0.41302(9) & 120(6) & 0.279(6) & 5.4 \\
		(0-300d)& $2f_{rot}$&0.8254(1) & 108(6) & 0.175(8) & 5.8 \\
		& $3f_{rot}$&1.2383(1) & 89(6) & 0.61(1) & 7.7 \\
		\midrule
		2 & $f_{rot}$ &0.41259(4) & 173(3) & 0.164(3) & 17.0 \\
		(300-589d)& $2f_{rot}$&0.82553(6) & 114(3) & 0.074(5) & 13.1 \\
		& $3f_{rot}$&1.23846(9) & 73(3) & 0.939(7) & 11.5 \\   
		\midrule
		3 & $f_{rot}$ &0.41267(3) & 262(4) & 0.730(2) & 23.4 \\
		(604-900d)& $2f_{rot}$&0.82551(7) & 105(4) & 0.938(6) & 10.9 \\
		& $3f_{rot}$&1.2384(1) & 65(4) & 0.298(9) & 9.4 \\   
		\midrule
		4 & $f_{rot}$ &0.41271(4) & 232(5) & 0.370(3) & 16.2 \\
		(900-1200d)& $2f_{rot}$&0.8254(1) & 92(5) & 0.667(9) & 7.2 \\
		& $3f_{rot}$&1.2381(2) & 46(5) & 0.53(2) & 4.3 \\   
		\midrule
		5 & $f_{rot}$ &0.41289(9) & 157(7) & 0.117(7) & 6.6 \\
		(1200-1461d)& $2f_{rot}$&0.8258(2) & 78(6) & 0.91(1) & 4.2 \\
		& $3f_{rot}$&1.2385(2) & 43(6) & 0.79(2) & 3.4 \\  
		\bottomrule
	\end{tabular}
	\label{tab:lowfrequency}
	\tablecomments{ Rotational frequency and its harmonics have been extracted using  Kepler LC data which are cut into 5 segments started from BJD2454964.0. The data quality in Segment2, 3 and 4 is better than in Segment 1 and 5.}
\end{table}
\begin{figure}
    \centering
    \includegraphics[width=1\linewidth]{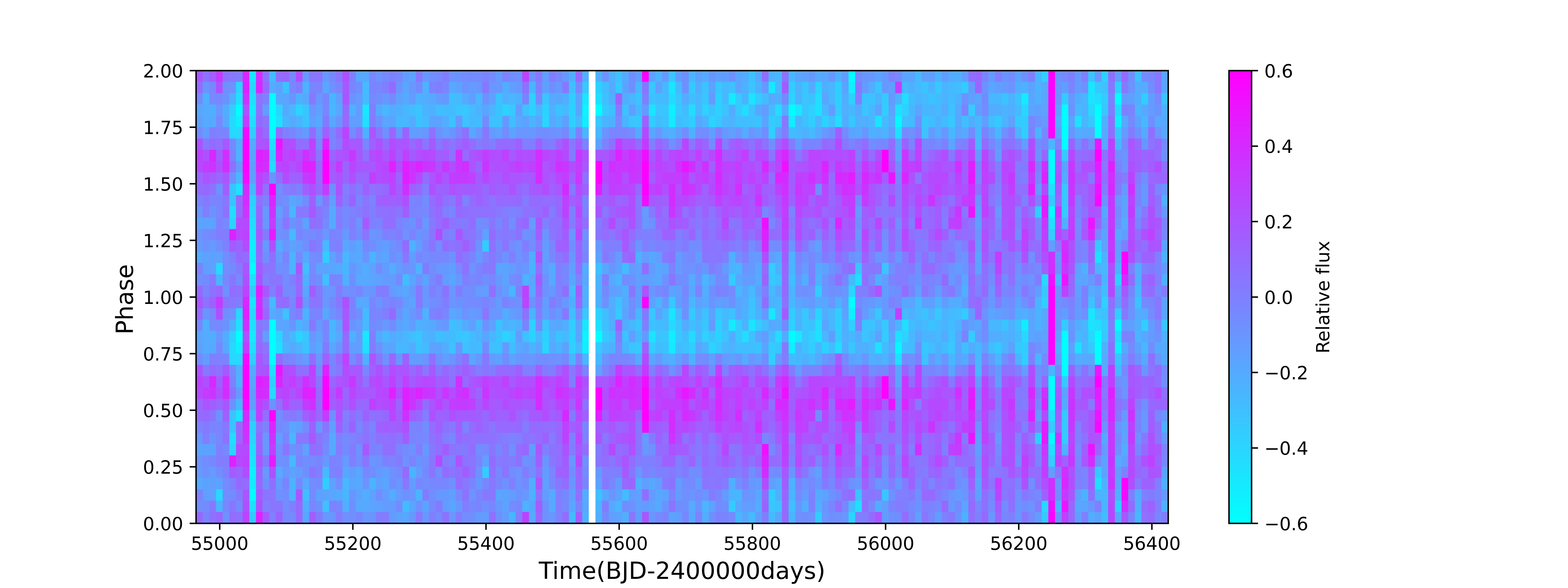}
    \caption{The phase evolution diagram over a time span of about four years for the residuals of the Kepler LC data. Pixel shades, from blue to purple, indicate the mean flux in each (time, phase) bin from low to high.  Each pixel represents 10 days in time-axis and 0.05 in phase-axis. The vertical white gap inside exists due to a gap of data.}
    \label{fig:pixel}
\end{figure}

If the co-rotational structure like star spots resulted in modulation frequencies, then their  evolution will also result in time-domiain variation in amplitude.  Two efforts have been made to figure this out. Firstly, we inspected TESS data in order to expand time span, however, the entire low frequency region disappear in amplitude spectra maybe because of the poor quality of our processed data. Another effort we made is to create a continuous evolution diagram by folding phase at $\nu=0.4126{\text{d}}^{-1}$, which is shown in Figure \ref{fig:pixel}. The residual data are folded twice for clarity. In Figure \ref{fig:pixel}, each row represents how the mean flux in 0.05 phase of data changes with time, each column indicates how mean flux in 10 days changes within two rotational periods. Every (time, phase) pixel contains the information of mean flux value. Considering the diagram as a whole, the brightness on the star's surface is uneven over a time span of about 1460 days. The first and the last segments show chaos due to imperfect data quality and low S/N. Nevertheless, the overall trend of the evolution of flux is still clear from left to right in time-axis, that high-flux area is gradually expanding in rotational period. The $f_{rot}$ we chose might be slightly imperfect and it will affect the slope of the boundary of transition, while the expansion is clear no matter  where the boundary is. It indicates that the star appears brighter in the rotational period along with time. If the hypothesis of rotation modulation really happens in the star, then spots on surface may be fading and their influential zone is decreasing.

Based on the assumption that $\nu=0.4126{\text{d}}^{-1}$ represents $f_{rot}$, we can infer the rotational velocity of KIC 10855535. The $R/R_{\odot}$ value can be derived respectively from KIC and TIC while without uncertainties. Using stellar parameters from a low-resolution spectrum (LRS, R=1800) obtained from the LAMOST (Large Sky Area Multi-Object Fiber Spectroscopic Telescope; \citealt{2012RAA....12.1197C}), we apply the empirical relation between$log R$ and $T_{\text{eff}} +\log g + [Fe/H] $ for estimating stellar radii \citep{2022A&A...663A.112M}. This yields $R/R_{\odot} = 1.754(0.1)$. Therefore, the upper limit of $v\sin i$ cannot exceed $v=36.5\pm 2.1 \text{km/s}$. Although we attempted to verify $v\sin i$ through spectroscopic observations, values below $100 \text{km/s}$ are undetectable in the case of Lamost LRS (R=1800) \citep{2024ApJS..271....4Z}. Our result still relies on the hypothesis that low frequencies originate from rotation modulation. Following this line of reasoning, the star could be classified as a slow rotator ($v\sin i\lesssim 50 \text{ km/s}$) \citep{2013AJ....145..132C,2024MNRAS.tmp.2167M}.

However, KIC 10855535 is very deffcient in metallicity according to two LAMOST LRS.
The discussion on this section founded on the inference that $f_8$ is $f_{rot}$, but a puzzle confuses us is that if the star is a very metal-poor A-type star, why star spots would be so powerful on it? We will leave the question here.

\section{Summary} \label{sec:summary}
By analyzing the Kepler data, KIC 10855535 has been confirmed to be a $\delta$ Sct star within a binary system. The amplitude spectrum is clean revealing only two independent frequencies: ${\text{F}}0$ and $f_{rot}$.  For F0, there is an equally spaced quintuplet centered around it, with a frequency splitting of $f_m=0.0243(1){\text{d}}^{-1}$. This equidistant splitting is the consequence of  frequency modulation caused by its companion. Orbital period can be calculated as its reciprocal at $411.5\pm1.7$ days.  

The main frequency F0 exhibits a quasi-sinusoidal modulation in its pulsation phase, is attributed to the light-time effect of the binary system. Maelstrom code has been applied to give a model of binary. Orbital parameters we obtained are consistent with previous research. To study the long-term variations of this star, TESS datasets were also utilized to derive time-delay caused by LTE. It portrays a simple scene that the star has been orbiting in the binary system with a period of $412.2(3)$ days during past two decades. After altering timestamps of light arrival time to the light emission time, quasi-sinusoidal phase modulation disappeared and was replaced by a parabolic change which indicates linear pulsation period growth with $\dot{P}/P\sim1.44\times10^{-7}\text{yr}^{-1}$. We think it is related to stellar evolution according to current knowledge.

Additionally, F0 is the only frequency that shows a stable amplitude growth. We have no idea about whether the modulation is transient. However, combined with the period growth meanwhile, it is possible that stable evolution brings some changes in the structure of star. Other speculations like non-linearity or mode coupling, stellar cycle are still not ruled out due to lack of additional evidence.

Regarding comb-like frequencies, simple and precise harmonics lying in the low-frequency range, which we referred as $f_{rot}$,  we examined its stability via prewhitening other high frequencies and checking amplitude variations in segments spanning hundreds of days. A flux pixel diagram has also been plotted to verify flux changes during each 'rotation period' and long term observation period. If the modulation is indeed produced by co-rotational structure like star spots on surface, then they are evolving with time and manifest variations on the flux. Based on this speculation, we deduced the upper limit of $v\sin i$ and found that, for a low-amplitude $\delta$ Sct star, it rotates relatively slowly. This slow rotation might be one of the reasons for its stable evolution.

In fact this star exhibits similar features to HADS, including a notably clean spectrum among $\delta $ Sct variables and a quite slow rotational velocity. However, its amplitude is not sufficiently high and its metallicity is too low to be classified as a HADS. \citet{2021AJ....162...48L} reported a double-mode low amplitude KIC 12602250 and highlighted its importance on understanding true features of HADS. Compared with it, KIC 10855535 has an even cleaner spectrum making it a prime candidate for detailed research to explore differences between HADS and normal $\delta$ Sct stars.  Combined all these features together: (a)low amplitude, (b)stable pulsation amplitude growth, (c)slow rotation and (d)clean spectrum, KIC 10855535 gives us an intuitive feeling that the star is elegant. The advantage of this kind of 'elegant' is that it avoids hardship in identifying pulsation modes and distraction from complicated interaction, thus allowing the underlying characteristics to be revealed more quickly and directly.

Due to the limited pulsational modes in KIC 10855535, further investigations, such as constructing oscillation models, are constrained. While high-resolution spectroscopic and long-term photometric observations are warranted, the faintness of the star poses challenges. Nonetheless, this case offers valuable insights into the relationship between stellar evolution and pulsation, potentially establishing new connections between typical $\delta$ Sct and HADS.

\begin{acknowledgments}

The authors would like to thank the reviewer's insightful comments, which have greatly improved this work. The work is supported by the National Key R$\&$D program of China for the Intergovernmental Scientific and Technological Innovation Cooperation Project under No. 2022YFE0126200, and Tian-shan Talent Training Program under No. 2023TSYCLJ0053. J.Z.L was supported by the Tianshan Talent Training Program  through the grant 2023TSYCCX0101. We sincerely thank the teams that have contributed to Kepler and TESS missions and have been operating and maintaining LAMOST. 

\end{acknowledgments}

\software{Astropy \citep{2013A&A...558A..33A,2018AJ....156..123A,2022ApJ...935..167A},
	Matplotlib \citep{thomas_a_caswell_2022_6513224},
	Lightkurve \citep{2018ascl.soft12013L,geert_barentsen_2021_4654522} } 

\bibliography{reference}

\end{document}